\documentclass[12pt]{article}

\usepackage[table,xcdraw]{xcolor}   
\usepackage[T1]{fontenc}
\usepackage[english]{babel}
\usepackage{lmodern}
\usepackage[sfdefault]{libertinus}  
\usepackage{newtxtext,newtxmath}    

\usepackage{amsmath}
\usepackage{amssymb}
\usepackage{amsthm}

\usepackage{graphicx}
\usepackage{epsfig}
\usepackage{algorithmicx}
\usepackage{algcompatible}

\usepackage{caption}
\usepackage{subcaption}
\usepackage{tikz}
\usepackage{smartdiagram}
\usetikzlibrary{shapes.geometric, arrows.meta, positioning}

\tikzstyle{block} = [rectangle, rounded corners, minimum width=3.8cm, minimum height=1cm, text centered, draw=blue!60, fill=blue!10, text width=4.2cm, font=\footnotesize]
\tikzstyle{arrow} = [thick,->,>=latex]

\usepackage{algorithm}
\usepackage{algpseudocode}

\usepackage{booktabs}
\usepackage{multirow}
\usepackage{longtable}
\usepackage{array}
\usepackage{colortbl}
\usepackage{adjustbox}

\usepackage{setspace}
\usepackage{titlesec}
\usepackage{indentfirst}
\usepackage{enumitem}
\usepackage{float}
\usepackage{here}
\usepackage{lipsum}
\usepackage{geometry}

\usepackage{hyperref}
\usepackage{cite}
\usepackage{url}

\titleformat{\section}[block]{\rmfamily\bfseries}{\thesection}{1em}{}

\titlespacing*{\section}{0pt}{0pt}{0pt}

\newcommand{\headingfont}{\rmfamily}

\begin{document}

\begin{center}
{\headingfont\fontsize{18}{22}\selectfont\textbf{Asymptotic Consistency and Generalization in Hybrid Algorithms of Regularized Selection and Nonlinear Learning}}
\end{center}

\begin{center}
\headingfont\textbf{Luciano Ribeiro Galvão}$^{1*}$ \\
\headingfont\textbf{Rafael de Andrade Moral}$^2$  \\[2ex]

$^1$ Department of Exact Sciences, University of S\~ao Paulo, Brazil\\
$^2$ Department of Mathematics and Statistics, Maynooth University, Ireland\\
$^*$ corresponding author: lucianogalvao@usp.br\\

\end{center}

\headingfont\textbf{\textit{Abstract}}

This study explores how different types of supervised models perform in the task of predicting and selecting relevant variables in high-dimensional contexts, especially when the data is very noisy. We analyzed three approaches: regularized models (Lasso, Ridge, and Elastic Net), black-box models (Random Forest, XGBoost, LightGBM, CatBoost, and H2O GBM), and hybrid Algorithms that combine both approaches: regularization with nonlinear algorithms. Using simulations based on the Friedman equation, we evaluated 23 algorithms using three complementary metrics: RMSE, Jaccard index, and recovery rate. The results revealed that, although black-box models excel in predictive accuracy, they lack interpretability and simplicity, essential factors in many real-world contexts. Regularized models, on the other hand, proved to be more sensitive to an excess of irrelevant variables. In this scenario, hybrid Algorithms stood out for their balance: they maintain good predictive performance, identify relevant variables more consistently, and offer greater robustness, especially as the sample size increases. Therefore, we recommend using this hybrid framework in market applications, where it is essential that the results make sense in a practical context and support decisions with confidence.

\begin{singlespacing}

\end{singlespacing}

\vspace{0,60cm}

\textit{{\headingfont\textbf{Key Words: Regularization; Black-box models; Variable selection; High dimensionality; RMSE; Jaccard; Model interpretation.}} }
\noindent

\vspace{0,60cm}

\section{\headingfont Introduction}

\begin{spacing}{1.5}
\noindent
Predictive models for high-dimensional problems pose a significant computational challenge. In many cases, the direct use of nonlinear algorithms to reduce the number of features can be computationally expensive. Although widely used, this framework can generate biases, overfitting (induced by heuristics), and limitations in methods based on variable importance, given that the predictive model itself influences the prioritization of selected variables \cite{guyon2003introduction, hastie2009elements}. The use of non-heuristic methods, such as regularization independent of the predictive model, has gained prominence in several fields to mitigate such biases, with hybrid Algorithms being a promising solution \cite{meinshausen2010stability}. However, these associated models require further theoretical development regarding their generalization capacity and asymptotic properties.

This work attempts to develop a theoretical framework that analyzes the selection consistency and generalization limits of these hybrid Algorithms, focusing on the combination of regularized methods such as Lasso, Ridge, and Elastic Net with widely used black-box algorithms, including Random Forest, XGBoost, LightGBM, CatBoost, and H2O GBM.

Regarding the selection step using regularized models, Lasso (\cite{zhao2006lasso, wainwright2009sharp}) is a selection and regularization method that promotes sparsity in the coefficients through the $\ell_1$ penalty. Lasso is consistent in recovering the true set of variables depends on conditions such as unrepresentability and strong signal strength, which are essential in high-dimensional scenarios. Ridge Regression (\cite{hoerl1970ridge}) applies the quadratic penalty ($\ell_2$) and does not perform variable selection, but contributes to greater stability and variance reduction in the presence of multicollinearity. Elastic Net \cite{zou2005elastic} combines the $\ell_1$ and $\ell_2$ penalties, offering a robust alternative to Lasso, which is especially effective in the presence of high correlations between predictors.

Black-box models are widely used in supervised learning applications due to their high predictive capacity and flexibility. Random Forest is based on bagging \cite{buhlmann2002analyzing}, a technique that reduces variance by combining decision trees generated on bootstrapped subsets. Boosting-based models, such as XGBoost, LightGBM, and CatBoost, use sequential additive learning to reduce bias and improve performance, while presenting challenges in terms of stability and overfitting \cite{hardt2016train}. H2O GBM offers a scalable implementation of boosting with automatic hyperparameter adjustments. The theory underlying these models includes interpretations such as adaptive neighbors for Random Forest \cite{lin2006random} and bias-variance analysis for more randomized ensembles such as Extremely Randomized Trees\cite{geurts2006extremely}.

Algorithmic stability plays a central role in theoretical analysis and is directly connected to the behavior of generalization errors, as discussed by Bousquet and Elisseeff (2002) \cite{bousquet2002stability}. In this context, the notion of expected risk consistency stands out. This is related to the Empirical Risk Minimization Principle (ERM), as presented by Vapnik \cite{vapnik1999nature}(1999). This principle establishes that a good model must present asymptotic convergence between the empirical risk observed in the data and the expected risk associated with the true function that generated the data. Such convergence is essential in supervised learning scenarios, as it ensures that, as the number of observations increases, the model's performance tends to approach optimality within the class of functions considered.

This property is particularly relevant in scenarios where predictive accuracy is more important than the exact recovery of relevant variables, but here it is used to demonstrate the assertive power of the proposed hybrid models. In this context, even without identifying all the true support, a model can be considered predictive consistent if its predictive ability remains robust as the number of observations increases. The loss function and the type of penalty also impact the bias and variance of the estimator, \cite{rosasco2004loss}.

To empirically evaluate the proposed framework, controlled simulations were performed based on the Friedman equation, Friedman (1991) \cite{friedman1991}. This equation is widely used in the literature to generate synthetic data with a known ground truth structure, including relevant predictor variables and noise components. The experimental design allows a clear distinction between the true model and the estimated model, enabling the evaluation of method performance in different dimensionality regimes ($n$ and $p$) and under different levels of noise and structural complexity of the data. Despite the vast literature on regularized methods and black-box models used individually, there is a lack of a theoretical framework that rigorously connects and grounds these steps in the hybrid context. This article aims to contribute to filling this gap by offering new theoretical constraints and conditions for the consistency and generalization of the proposed hybrid pipeline.

\section{Methodology}\label{methodology}

Variable selection in high-dimensional scenarios requires formal criteria that guarantee the theoretical validity of the methods used, especially when seeking to ensure stable performance in large samples. In this section, we discuss two fundamental concepts in this context: \textit{selection consistency}, which ensures the recovery of true support, and \textit{expected risk consistency}, which ensures the convergence of the prediction error to the optimal risk of the generating model.

These properties are essential for the asymptotic robustness of methods that combine regularization and nonlinear modeling, such as the hybrid frameworks proposed in this study. The detailed structure of the experiment is presented in Algorithm \ref{pseudo_cod1}.

\begin{algorithm}[H]
\caption{Simulation and Evaluation Framework with Hybrid Algorithms}
\label{pseudo_cod1}
\small
\begin{algorithmic}[1]
\Require Dataset $\mathcal{D} = \{(x_i, y_i)\}_{i=1}^n$, where $x_i \in \mathbb{R}^p$, $y_i \in \mathbb{R}$; sample sizes $n \in \{50, 100, 200, 500, 1000\}$; number of predictors $p \in \{5, 10, 50, 100 \}$; replicates $n_{\text{sim}}$; performance metrics: RMSE, Jaccard index, Recovery.
\For{sim = 1 to $n_{\text{sim}}$}
    \State Generate synthetic data:
    \[
    x_{ij} \sim \mathcal{U}(0,1), \quad y_i = 10 \sin(\pi x_{i1} x_{i2}) + 20(x_{i3} - 0.5)^2 + 10x_{i4} + 5x_{i5} + \varepsilon_i,\quad \varepsilon_i \sim \mathcal{N}(0,1)
    \]
    \For{each model in \{Ridge, Lasso, Elastic Net\}}
        \State Set $\alpha = 0$ (Ridge), $\alpha = 1$ (Lasso), $0 < \alpha < 1$ (Elastic Net).
        \State Fit \texttt{cv.glmnet} with family = "gaussian":
        \begin{itemize}
            \item Use MSE to select $\lambda_{\min}$.
            \item Extract coefficients $\hat{\beta}_j$ at $\lambda_{\min}$.
        \end{itemize}
        \State Rank variables by $|\hat{\beta}_j|$, forming $\hat{S}_{\text{ranked}}$.
        \For{m = 1 to min(10, p - 1)}
            \State Fit OLS on top-m variables from $\hat{S}_{\text{ranked}}$.
            \State Evaluate RMSE using k-fold CV.
        \EndFor
        \State Select $m^*$ with minimal RMSE and define $\hat{S} = \hat{S}_{\text{ranked}}[1:m^*]$.
        \State Compute:
        \[
        \text{RMSE}_{\text{reg}}, \quad
        \text{Jaccard} = \frac{|\hat{S} \cap S|}{|\hat{S} \cup S|}, \quad
        \text{Recovery} = \frac{|\hat{S} \cap S|}{|S|}
        \]
    \EndFor

    \For{each black-box model in \{RF, XGBoost, LightGBM, CatBoost, H2O GBM\}}
        \State Fit model using all predictors to extract feature importance.
        \State Rank variables by importance and form $\hat{S}_{\text{imp}}$ with up to 10 variables.
        \For{m = 1 to min(10, p - 1)}
            \State Evaluate RMSE via k-fold CV on top-m variables from $\hat{S}_{\text{imp}}$.
        \EndFor
        \State Select $m^*$ with minimal RMSE and define $\hat{S}_{\text{bb}} = \hat{S}_{\text{imp}}[1:m^*]$.
        \State Compute $\text{RMSE}_{\text{bb}}$, Jaccard and Recovery using $\hat{S}_{\text{bb}}$.
    \EndFor

    \For{each combination of black-box and regularized subset}
        \State Apply black-box model on $\hat{S}$ from regularized model (hybrid).
        \State Compute RMSE, Jaccard and Recovery for hybrid.
    \EndFor
\EndFor
\State \textbf{Return} all stored metrics (RMSE, Jaccard, Recovery) across simulations.
\end{algorithmic}
\vspace{0.3em}
\end{algorithm}

Next, we formalize some of the main definitions in the general context of the regression problem according to the proposed framework.

\paragraph{Definition: Asymptotic Consistency}  To characterize methods that correctly identify the truly relevant variables in the asymptotic limit, following
\cite{buhlmann2011statistics} and \cite{shalev2014understanding}, one can define $S \subset \{1, \dots, p\}$ as the truly relevant set of variables and $\hat S$ as the set estimated by the model. We say that a variable selection method is \emph{selection-consistent} if
\begin{equation}
P\left( \hat S = S \right) \xrightarrow[n \to \infty]{} 1.
\end{equation}

Furthermore, we say that the predictive estimator $\hat f$ is \emph{predictive consistent} if
\begin{equation}
\mathbb{E}\left[ \left( f(\mathbf{X}) - \hat f(\mathbf{X}) \right)^2 \right] \xrightarrow[n \to \infty]{}  0,
\end{equation}
where $f(\mathbf{X})$ is the truth function of the generative model.

\paragraph{Definition: Generalization}

Let $R(\hat f)$ be the expected risk in the population:
\begin{equation}
R(\hat f) = \mathbb{E} {(Y, \mathbf{X})} \left[ L(Y, \hat f(\mathbf{X})) \right],
\end{equation}
where $L(\cdot)$ is an appropriate loss function (e.g., squared error, deviance, log-loss). The model generalizes well if
\begin{equation}
R(\hat f) - R(f^*) \xrightarrow[n \to \infty]{} 0,
\end{equation}
where $f^*$ is the minimizer of the expected risk.

\paragraph{Consistency of the Hybrid Framework.}
Suppose that the true model has a sparse structure with support \( S \subset \{1, \dots, p\} \), and that the regularized selection step satisfies the conditions of unrepresentability and signal strength, ensuring \( \mathbb{P}(\hat{S} = S) \to 1 \) when \( n \to \infty \). Further suppose that the predictive model fitted over the selected subspace (\textit{black-box} model) is consistently predictive in the sense of \( R(\hat{f}) \to R(f^*) \). Under these assumptions, the resulting hybrid framework would also be consistent in both support recovery and predictive risk.

This conjecture expresses a principle proposed in this work based on two widely recognized theoretical pillars: the consistency of selection via regularization and the predictive consistency of models fitted on relevant subsets~\cite{scornet2015consistency}.

\subsection{Experimental Study}

To empirically validate our findings, we chose to carry out simulation studies. The simulations were based on the Friedman model, widely used in the literature to evaluate variable selection methods in nonlinear and high-dimensional scenarios. What is considered a true model corresponds to the case in which the predictor vector \( \mathbf{x}_i \in \mathbb{R}^p \) has real support \( S = \{1, 2, 3, 4, 5\} \), meaning that only the first five variables affect the response.

In this context, the model with additive noise is defined by:
\begin{equation}
y_i = f(\mathbf{x}_i) + \varepsilon_i = 10 \sin(\pi x_{i1} x_{i2}) + 20 (x_{i3} - 0.5)^2 + 10 x_{i4} + 5 x_{i5} + \varepsilon_i,
\end{equation}
where \( \varepsilon_i \sim \mathcal{N}(0, 1) \) represents the error term, and each covariate \( x_{ij} \overset{\text{i.i.d.}}{\sim} \mathcal{U}(0, 1),\ \mbox{for}\ j = 1, \dots, p \).

Although only the first five variables are present in the response generating function \( f(\mathbf{x}_i) \), here considered the true support, the vector \( \mathbf{x}_i \) ranges from \( p = 5 \) to \( p = 100 \) covariates, of which \( p - 5 \) are irrelevant and represent the stochastic noise in the model.

To evaluate the methods' ability to recover the true set of relevant variables, we also consider a deterministic version of the model, that is, without additive noise, defined by:

\begin{equation}
y_i = f(\mathbf{x}_i) = 10 \sin(\pi x_{i1} x_{i2}) + 20 (x_{i3} - 0.5)^2 + 10 x_{i4} + 5 x_{i5}.
\end{equation}

This noise-free model allows for the isolated measurement of the exact recovery capacity of variable selection methods, as it eliminates the interference of random error.

\subsection{Models Evaluated}

In this study, three main groups of models were considered. The first group comprises the \textit{regularized} models: Ridge (with penalty \(\alpha = 0\)), Lasso (\(\alpha = 1\)), and Elastic Net (\(\alpha = 0.5\)), fitted using the \textit{glmnet} package \cite{friedman2010glmnet}. Variable selection in these models was performed based on the support of the non-zero coefficients obtained at the penalty point \(\lambda_{\min}\), chosen via cross-validation.

The second group includes \textit{black-box} models, which include Random Forest \cite{breiman2001randomforest}, XGBoost \cite{chen2016xgboost}, LightGBM \cite{ke2017lightgbm}, CatBoost \cite{prokhorenkova2018catboost}, and H2O GBM. These algorithms were fitted on all available predictors, without a prior selection step.

Finally, the third group consists of \textit{hybrid} Algorithms, in which each black-box model was fitted exclusively on the subset of variables previously selected by one of the regularized models, seeking to combine parsimony with predictive flexibility.

\subsection{Simulations}\label{subsec:simulate}

This study evaluated the performance of regularized, pure, and hybrid black-box algorithms in twelve distinct scenarios, defined by combinations of five sample sizes (n = 50, 100, 200, 500, and 1,000) and four dimensionality levels (p = 5, 10, 50, and 100). For each scenario, ten simulated datasets were generated independently and replicated for each of the three proposed metrics (RMSE, Jaccard, and Recovery).

For each simulated dataset, 23 algorithms were applied: three regularized models (Ridge, Lasso, and Elastic Net), five pure black-box models (Random Forest, XGBoost, LightGBM, CatBoost, and H2O GBM), and 15 hybrid algorithms resulting from the combination of regularization-based and black-box algorithms.

Thus, each scenario involved \(230\) runs \((10 \text{ datasets} \times 23 \text{ algorithms})\), 
replicated for 3 metrics and 20 associations of \(n\) and \(p\), totaling 
\(20 \times 230 \times 3 = 13{,}800\) runs. 
Table~\ref{tab:cenarios_simulacao} details the scenarios considered and the composition of the models evaluated by type. 
In each run, the data were randomly partitioned into training (80\%) and testing (20\%) sets to calculate the validation metrics.

\begin{table}[H]
\centering
\caption{Simulation parameters, number of executions per scenario, and composition of model types.}
\label{tab:cenarios_simulacao}
\begin{tabular}{ccccc}
\toprule
\textbf{\(n\)} & \textbf{\(p\)} & \textbf{Scenario} & \textbf{Model} & \textbf{Quantity} \\
\midrule
\multirow{3}{*}{50 | 100 | 200 | 500 | 1000} & \multirow{3}{*}{5} & \multirow{3}{*}{230} & Regularized & 3 \\ 
& & & Pure black-box & 5 \\ 
& & & Hybrids (Reg × BB) & 15 \\
\midrule
\multirow{3}{*}{50 | 100 | 200 | 500 | 1000} & \multirow{3}{*}{10} & \multirow{3}{*}{230} & Regularized & 3 \\ 
& & & Pure black-box & 5 \\ 
& & & Hybrids (Reg × BB) & 15 \\
\midrule
\multirow{3}{*}{50 | 100 | 200 | 500 | 1000} & \multirow{3}{*}{50} & \multirow{3}{*}{230} & Regularized & 3 \\ 
& & & Pure black-box & 5 \\ 
& & & Hybrids (Reg × BB) & 15 \\
\midrule
\multirow{3}{*}{50 | 100 | 200 | 500 | 1000} & \multirow{3}{*}{100} & \multirow{3}{*}{230} & Regularized & 3 \\
& & & Pure black-box & 5 \\
& & & Hybrids (Reg × BB) & 15 \\
\bottomrule
\end{tabular}
\end{table}

The regularized models then fitted using the penalty corresponding to the optimal value \(\lambda_{\text{min}}\), obtained by cross-validation using the lowest RMSE criterion per feature subset. Based on the set of variables selected by these models, the hybrid algorithms then utilised the black-box methods based exclusively to the selected predictors. For comparison purposes, the pure black-box models were also fitted to the data, considering all available predictors, regardless of relevance. The final performance of the models was then evaluated based on the RMSE (Root Mean Squared Error), Jaccard index, and Recovery metrics.

To ensure comparability between the methods, the black-box algorithms were trained with 100 trees or iterations. In particular, the models based on the H2O package \cite{h2o_package} were initialized with full parallelism to optimize computational performance.

In Section ~\ref{sec:results}, we focus on the values of \(p = \{50\}\) and \(n = \{200, 500, 1000\}\) and their main results for the purposes of brevity and clarity. However, the full range evaluated is presented in section \ref{AnexoI}, in table \ref{tab:hybrid_models_full}.

\subsection{Evaluation Metrics}

Model performance was evaluated using three metrics. The first is RMSE (Root Mean Squared Error), which quantifies the mean squared error of predictions on the test set and is widely used as a measure of predictive accuracy.

The other two metrics refer to the quality of variable selection: the Jaccard index \cite{jaccard1901distribution}, defined as
\[
J = \frac{|S \cap \hat{S}|}{|S \cup \hat{S}|},
\]
and the recovery rate \cite{wasserman2009high}, given by
\[
R = \frac{|S \cap \hat{S}|}{|S|},
\]
where \(S\) represents the true set of relevant variables and \(\hat{S}\) the set estimated by the model. Both are widely adopted in the literature to assess the consistency of variable selection.

\subsection{Computer Implementation}

The simulations were implemented in the \textit{R}~\cite{R} software, using the following packages:
\textit{glmnet} (regularized models) \cite{glmnet},
\textit{randomForest} (Random Forest) \cite{randomForest},
\textit{xgboost} \cite{xgboost},
\textit{lightgbm} \cite{lightgbm},
\textit{catboost} \cite{catboost},
\textit{h2o} (with parallelism enabled for H2O GBM) \cite{h2o_package},
\textit{dplyr} \cite{dplyr},
\textit{tibble} \cite{tibble},
\textit{tidyr}\cite{tidyr},
\textit{Metrics}\cite{Metrics}
\textit{purrr}\cite{purrr},
\textit{doParallel}\cite{doParallel}, and
\textit{foreach} (for parallelization of simulations).

The results were saved in \texttt{.csv} and \texttt{.RData} files for reproducibility. Graphical visualizations were generated with the \textit{ggplot2} and \textit{viridis} packages. The source code is publicly available \cite{Galvao2025_b}.

\vspace{0.65cm}

\section{Results}\label{sec:results}

In this section, we present the results obtained in the simulation studies. The evaluation compared different modeling strategies, including regularized approaches, black-box models, and hybrid Algorithms that combine selection via regularization with nonlinear predictors. The evaluation metrics considered were the root mean square error (RMSE), the Jaccard index (for similarity between selected variables and the true set), and the recovery rate of relevant variables. The models in the subsections are analyzed segmented, allowing us to evaluate the relative performance of each class of approach under different complexity and generalization criteria.

\subsection{Pure Black-Box Models}

Table \ref{tab:blackbox_models_complete_p50} presents the results of the main black-box models in their pure form, denoted as \textit{Full} because they use the same algorithm for both variable selection and prediction, without combining them with regularized methods. Five algorithms are considered: CatBoost, H2O GBM, LightGBM, Random Forest, and XGBoost. The comparison includes three metrics at different sample sizes: root mean square error (RMSE), Jaccard index, and recovery rate.

\begin{table}[H]
\centering
\caption{Pure black-box models with 90\% noise ($p = 50$). Values represent the mean (standard deviation) across 10 simulations. Bold highlights the best model per metric and sample size.}
\label{tab:blackbox_models_complete_p50}
\begin{tabular}{llccc}
\toprule
\textbf{Model} & \textbf{Metric} & \textbf{$n = 200$} & \textbf{$n = 500$} & \textbf{$n = 1000$} \\
\midrule
RF        & RMSE     & 2.19 (0.73) & 2.16 (0.05) & 1.75 (0.52) \\
XGBoost   & RMSE     & 2.24 (0.15) & 1.77 (0.53) & 1.69 (0.05) \\
LightGBM  & RMSE     & \textbf{1.86 (0.56)} & 1.75 (0.05) & 1.42 (0.42) \\
CatBoost  & RMSE     & 1.87 (0.09) & \textbf{1.56 (0.04)} & \textbf{1.10 (0.48)} \\
H2O GBM   & RMSE     & 2.02 (0.09) & 1.80 (0.04) & 1.45 (0.43) \\
\midrule
RF        & Jaccard  & 0.83 (0.00) & 0.83 (0.00) & 0.83 (0.00) \\
XGBoost   & Jaccard  & 0.91 (0.12) & 0.88 (0.15) & \textbf{0.97} (0.07) \\
LightGBM  & Jaccard  & 0.76 (0.17) & 0.76 (0.15) & 0.89 (0.14) \\
CatBoost  & Jaccard  & \textbf{0.91 (0.12)} & \textbf{0.95 (0.08)} & \textbf{0.97} (0.07) \\
H2O GBM   & Jaccard  & 0.89 (0.16) & 0.85 (0.15) & 0.86 (0.18) \\
\midrule
RF        & Recovery & 1.00 (0.00) & 1.00 (0.00) & 1.00 (0.00) \\
XGBoost   & Recovery & 0.98 (0.06) & 1.00 (0.00) & 1.00 (0.00) \\
LightGBM  & Recovery & 1.00 (0.00) & 1.00 (0.00) & 1.00 (0.00) \\
CatBoost  & Recovery & 1.00 (0.00) & 1.00 (0.00) & 1.00 (0.00) \\
H2O GBM   & Recovery & 1.00 (0.00) & 1.00 (0.00) & 1.00 (0.00) \\
\bottomrule
\end{tabular}
\end{table}

The results demonstrated that black-box models stand out for their high predictive power, with consistently low RMSE values, as already cited in the literature and therefore expected in this work. The LGBM model obtained the lowest RMSE (\(1.86\)) when taking a sample of size \(200\) but its Jaccard index is relatively low compared to the others, obtaining the lowest metric for two of the three samples presented (\(n=200\); \(J = 0.76\) and \(n=500\); \(J=0.76\)). For larger samples (\(n=500\) and \(n=1000\)) Catboost performed substantially better than the others, with root mean square error values of $1.56$ and $1.10$. The model also presents high Jaccard and Recovery values, even for smaller sample sizes (\(n=200\) and \(n=500\)), indicating stability in identifying relevant variables. In terms of overlap with the true support, all models demonstrated excellent performance overall, with average recovery rates equal to or close to $1$ in all scenarios, with the exception of XGBoost, which recorded $0.98$ in smaller samples.

Although effective in predictive terms, pure black-box models do not promote, a priori, parsimony and offer little transparency in variable selection. This represents a significant obstacle in contexts that require interpretation or validation of the predictors used.

\subsection{Pure Regularized Models}

Table \ref{tab:regularized_models_p50} presents the performance of penalized models fitted directly to simulated data with high spurious dimensionality (p = 50). These include Ridge, Lasso, and Elastic Net, methods that seek to reconcile prediction and selection through coefficient regularization.

\begin{table}[H]
\centering
\caption{Pure regularized models with 90\% noise ($p = 50$). Values represent the mean (standard deviation) across 10 simulations. Bold indicates the best value per metric and sample size.}
\label{tab:regularized_models_p50}
\begin{tabular}{llccc}
\toprule
\textbf{Model} & \textbf{Metric} & \textbf{$n = 200$} & \textbf{$n = 500$} & \textbf{$n = 1000$} \\
\midrule
Ridge      & RMSE & 2.54 (0.16) & 2.65 (0.08) & 1.92 (1.08) \\
Lasso      & RMSE & 2.54 (0.16) & 2.65 (0.08) & \textbf{2.64 (0.05)} \\
ElasticNet & RMSE & \textbf{2.54 (0.17)} & \textbf{1.92 (1.08)} & 2.16 (0.95) \\
\midrule
Ridge      & Jaccard & 0.50 (0.13) & 0.44 (0.06) & 0.43 (0.07) \\
Lasso      & Jaccard & \textbf{0.50 (0.13)} & \textbf{0.45 (0.06)} & \textbf{0.46 (0.06)} \\
ElasticNet & Jaccard & 0.45 (0.09) & 0.45 (0.07) & 0.46 (0.05) \\
\midrule
Ridge      & Recovery & 0.84 (0.08) & 0.82 (0.06) & 0.82 (0.06) \\
Lasso      & Recovery & 0.84 (0.08) & 0.82 (0.06) & 0.80 (0.00) \\
ElasticNet & Recovery & \textbf{0.84 (0.08)} & \textbf{0.84 (0.08)} & \textbf{0.82 (0.06)} \\
\bottomrule
\end{tabular}
\end{table}

The regularized models exhibit good predictive performance, with relatively high RMSE values. Although they are still higher when compared to the table \ref{tab:blackbox_models_complete_p50} , there is a clear trend of improvement with increasing n, indicating that these methods benefit from greater sample information. 

The LASSO model with \(n=500\) and Ridge with \(n=1000\) stand out, both with values close to, and in some cases even higher than, the BlackBox models \(RMSE = 1.92\). Regarding variable selection, the recovery rate remains robust and stable ($R \approx 0.84$), revealing a good ability to identify truly relevant variables. Jaccard values performed poorly, suggesting that the models still select a considerable number of irrelevant variables, which can compromise parsimony and increase the risk of overfitting, especially in contexts with high noise.

\subsection{Hybrid Algorithms - Regularized Black boxes}

\begin{longtable}{llccc}
\caption{Hybrid regularized black-box models with $p = 50$. Values represent the mean (standard deviation) across 10 simulations. Bold indicates the best value per metric and sample size.}
\label{tab:hybrid_models_p50} \\
\toprule
\textbf{Model} & \textbf{Metric} & \textbf{$n = 200$} & \textbf{$n = 500$} & \textbf{$n = 1000$} \\
\midrule
\endfirsthead

\multicolumn{5}{c}{{\bfseries Table \thetable\ -- continued from previous page}} \\
\toprule
\textbf{Model} & \textbf{Metric} & \textbf{$n = 200$} & \textbf{$n = 500$} & \textbf{$n = 1000$} \\
\midrule
\endhead

\midrule \multicolumn{5}{r}{{Continued on next page}} \\
\endfoot

\bottomrule
\endlastfoot
CatBoost EN    & RMSE & 2.22 (0.70) & 1.85 (0.85) & \textbf{1.59 (0.91)} \\
CatBoost Lasso & RMSE & 2.25 (0.70) & 2.29 (0.19) & 1.99 (0.59) \\
CatBoost Ridge & RMSE & 2.19 (0.68) & 2.07 (0.65) & 1.89 (0.60) \\
H2OGBM EN      & RMSE & 2.20 (0.69) & \textbf{1.66 (0.96)} & 2.09 (0.16) \\
H2OGBM Lasso   & RMSE & 2.27 (0.71) & 2.28 (0.11) & 1.94 (0.58) \\
H2OGBM Ridge   & RMSE & 2.04 (0.93) & 2.05 (0.68) & 2.10 (0.16) \\
LightGBM EN    & RMSE & 2.01 (0.91) & 2.00 (0.61) & 2.10 (0.17) \\
LightGBM Lasso & RMSE & \textbf{2.00 (0.91)} & 1.85 (0.82) & 1.76 (0.77) \\
LightGBM Ridge & RMSE & 2.25 (0.71) & 2.06 (0.62) & 1.90 (0.58) \\
XGBoost EN     & RMSE & 2.17 (0.97) & 2.43 (0.15) & 2.06 (0.62) \\
XGBoost Lasso  & RMSE & 2.16 (1.02) & 2.25 (0.68) & 2.11 (0.63) \\
XGBoost Ridge  & RMSE & 2.67 (0.19) & 1.78 (1.00) & 2.05 (0.62) \\
RF EN          & RMSE & 2.53 (0.76) & 2.28 (0.76) & 2.37 (0.09) \\
RF Lasso       & RMSE & 2.50 (0.76) & 1.85 (1.10) & 2.37 (0.05) \\
RF Ridge       & RMSE & 2.50 (0.75) & 2.31 (0.70) & 2.35 (0.11) \\
\midrule
CatBoost EN    & Jaccard & 0.45 (0.09) & 0.45 (0.07) & 0.46 (0.05) \\
CatBoost Lasso & Jaccard & 0.50 (0.13) & 0.45 (0.06) & 0.46 (0.06) \\
CatBoost Ridge & Jaccard & 0.50 (0.13) & 0.44 (0.06) & 0.43 (0.07) \\
H2OGBM EN      & Jaccard & 0.45 (0.09) & 0.45 (0.07) & 0.46 (0.05) \\
H2OGBM Lasso   & Jaccard & 0.50 (0.13) & 0.45 (0.06) & 0.46 (0.06) \\
H2OGBM Ridge   & Jaccard & 0.44 (0.06) & 0.44 (0.06) & 0.43 (0.07) \\
LightGBM EN    & Jaccard & 0.45 (0.09) & 0.45 (0.07) & 0.46 (0.05) \\
LightGBM Lasso & Jaccard & 0.50 (0.13) & 0.45 (0.06) & 0.46 (0.06) \\
LightGBM Ridge & Jaccard & 0.45 (0.09) & 0.45 (0.07) & 0.46 (0.05) \\
XGBoost EN     & Jaccard & 0.44 (0.06) & 0.44 (0.06) & 0.43 (0.07) \\
XGBoost Lasso  & Jaccard & 0.44 (0.06) & 0.44 (0.06) & 0.43 (0.07) \\
XGBoost Ridge  & Jaccard & 0.44 (0.06) & 0.44 (0.06) & 0.43 (0.07) \\
RF EN          & Jaccard & 0.45 (0.09) & 0.45 (0.07) & 0.46 (0.05) \\
RF Lasso       & Jaccard & 0.50 (0.13) & 0.45 (0.06) & 0.46 (0.06) \\
RF Ridge       & Jaccard & 0.44 (0.06) & 0.44 (0.06) & 0.43 (0.07) \\
\midrule
CatBoost EN    & Recovery & 0.84 (0.08) & 0.84 (0.08) & 0.82 (0.06) \\
CatBoost Lasso & Recovery & 0.84 (0.08) & 0.82 (0.06) & 0.80 (0.00) \\
CatBoost Ridge & Recovery & 0.84 (0.08) & 0.82 (0.06) & 0.82 (0.06) \\
H2OGBM EN      & Recovery & 0.84 (0.08) & 0.84 (0.08) & 0.82 (0.06) \\
H2OGBM Lasso   & Recovery & 0.84 (0.08) & 0.82 (0.06) & 0.80 (0.00) \\
H2OGBM Ridge   & Recovery & 0.84 (0.08) & 0.82 (0.06) & 0.82 (0.06) \\
LightGBM EN    & Recovery & 0.84 (0.08) & 0.84 (0.08) & 0.82 (0.06) \\
LightGBM Lasso & Recovery & 0.84 (0.08) & 0.84 (0.08) & 0.82 (0.06) \\
LightGBM Ridge & Recovery & 0.84 (0.08) & 0.82 (0.06) & 0.82 (0.06) \\
XGBoost EN     & Recovery & 0.84 (0.08) & 0.82 (0.06) & 0.82 (0.06) \\
XGBoost Lasso  & Recovery & 0.84 (0.08) & 0.82 (0.06) & 0.82 (0.06) \\
XGBoost Ridge  & Recovery & 0.84 (0.08) & 0.82 (0.06) & 0.82 (0.06) \\
RF EN          & Recovery & 0.84 (0.08) & 0.82 (0.06) & 0.82 (0.06) \\
RF Lasso       & Recovery & 0.84 (0.08) & 0.82 (0.06) & 0.82 (0.06) \\
RF Ridge       & Recovery & 0.84 (0.08) & 0.82 (0.06) & 0.82 (0.06) \\

\end{longtable}

Table \ref{tab:hybrid_models_p50} shows that the hybrid algorithms present an adequate balance between predictive performance and robustness in selecting relevant variables. The RMSE result of the algorithm that combined the CatBoost and Ridge models, for example, obtained a root mean square error that surpasses in predictive power several of the pure Black Box models (\(RMSE = 1.59\)) and practically all regularized models alone, when we take $n = 1000$. Regarding the true support recovery indices, it is among the highest ((\(R = 82\))  and has a value within the averages observed for the others (\(J = 0.46\)).

The algorithm that combined H2OGBM with Elasticnet obtained impressive results for a sample size of $n = 500$, as did Catboost with Elasticnet. In these simulations, it outperformed four of the five pure nonlinear algorithm models tested in predictive power, obtaining an RMSE of $1.66$. In the Jaccard and Recovery index evaluation of this combined algorithm, H2OGBM with EN performed on par with the others, with a notable $R=0.84$, an indicator similar to the most accurate models in recovering the support set.

These results reinforce the value of Hybrid Algorithms in the practical environment, as they combine parsimony in selection with competitive performance in prediction, especially in scenarios with larger dimensions and samples.

\subsection{Comparative Analysis and Implications}

In this section, we observe how black-box models excel in predictive accuracy (lower RMSE) but lack native interpretability, while regularized models offer good interpretability but present limitations under high spurious dimensionality with high RMSEs. Hybrid algorithms occupy an intermediate position, combining good predictive performance and selection stability, especially as the sample size increases. As demonstrated in section \ref{sec:results}, in some of these scenarios, some of the hybrid algorithms, in addition to being more explainable, outperform some of the black-box models.

\begin{figure}[H]
\centering
\includegraphics[width=0.99\linewidth]{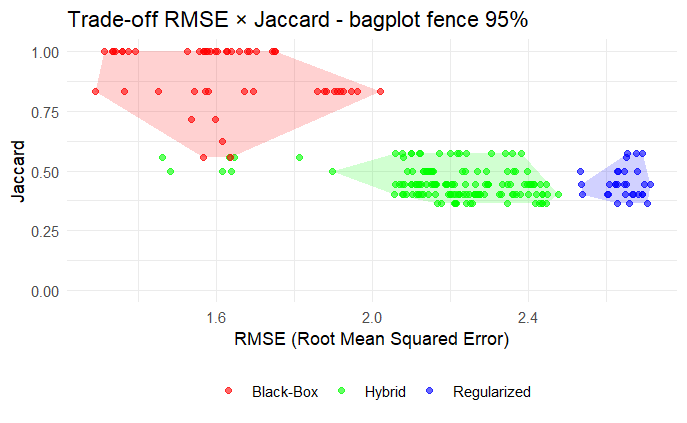}
\caption{Trade-off between predictive accuracy (RMSE) and selection consistency (Jaccard index) across the simulated models. Each dot corresponds to a simulation outcome, grouped into Black-Box, Hybrid, and Regularized classes. Shaded regions represent the robust 95\% confidence envelopes (bagplot fences), illustrating the empirical variability and stability of each class while respecting the bounded range of the Jaccard index.}
\label{fig:jac_rmse}
\end{figure}

Figure ~\ref{fig:jac_rmse} demonstrates, via Tukey's depth-based confidence regions \cite{rousseeuw1999bagplot}, this intermediate position between accuracy and consistency occupied by hybrid algorithms.

In \ref{fig:jac_rmse} , it shows the trade-off between accuracy and selection consistency among model classes. Black-box models obtained the lowest RMSEs in most cases, but with wider Jaccard indices compared to the others, demonstrating lower stability. Regularized models presented the opposite pattern, albeit with high RMSE values and low Jaccard indices. In the same image, it can be seen that hybrid algorithms occupy an intermediate position, balancing predictive accuracy with good recovery power of the real set, a relevant factor for scenarios with many variables.

\begin{figure}[H]
\centering
\includegraphics[width=0.99\linewidth]{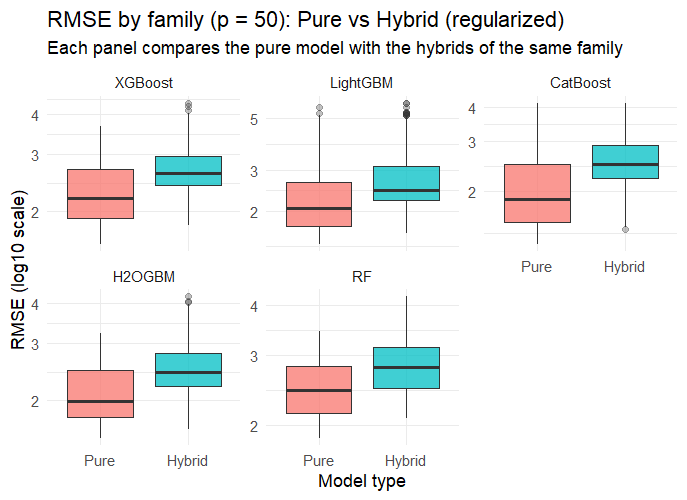}
\caption{Distribution of RMSE for each boosting and bagging family ($p = 50$). 
Each panel contrasts the pure version of the algorithm with its hybrid counterpart 
(Ridge/Lasso/ElasticNet for variable selection). Overlapping interquartile ranges 
indicate that hybrids only outperform the pure models in specific scenarios, depending 
on sample size $n$ and family.}
\label{fig:bp_family}
\end{figure}

When we evaluate the specific cases of associations for $p$ and $n$, we observe several scenarios where hybrid algorithms perform equally or better than pure nonlinear ones. In Figure \ref{fig:bp_family}, it is possible to observe that when segmenting into families of opaque models (Bagging and Boosting), there is a significant overlap in the interquartile ranges, suggesting that, in general, these points observed in specific scenarios are relevant, and there is no systematic superiority.

In summary, hybrid algorithms emerge as the most balanced solution for regression scenarios with high spurious dimensionality. They combine generalization, selection stability, and competitive performance. These findings are particularly relevant for applications in applied sciences and business environments, where accuracy and interpretability must go hand in hand.

\section{Conclusion}
\label{sec:conclusion}

Simulation results across different scenarios indicate that pure black-box models remain the most effective in terms of prediction error, but fail to promote parsimony and transparency. Conversely, pure regularized models exhibit moderate stability in recovering relevant variables, but present limitations in predictive accuracy, with systematically higher RMSE values and moderate Jaccard indices.

The hybrid algorithms demonstrated good performance when combining predictive accuracy with parsimony in variable selection. The CatBoost-ElasticNet and H2OGBM-ElasticNet combinations stood out, achieving low prediction errors and consistently identifying truly relevant variables. Our findings suggest that the balance between the \(l1\) and \(l2\) penalties of EN favors this combination of regularization and opaque models. It was observed that, as the sample size \(n\) increases, there is a clear improvement in the recovery and Jaccard metrics for these hybrid algorithms, indicating greater stability and accuracy in support reconstruction. This characteristic was observed even under conditions with a high number of spurious variables.

Our results reinforce the potential of hybrid algorithms in applications in the production sector and regulated environments where there is a strong need for transparency. These algorithms are especially useful in contexts where high dimensionality is a critical and unavoidable factor for business needs. It also applies to scenarios where interpretability of results is as important as predictive and computational performance. Therefore, by combining the selectivity of regularized methods with the flexibility of nonlinear algorithms, these approaches become robust and reliable solutions to real-world problems.

A possible future advancement is the creation of balanced continuous penalties integrated directly into the loss function of black-box models such as CatBoost or LGBM, which have demonstrated good penalty-associated performance, eliminating the need for external filtering steps.

\end{spacing}

\bibliographystyle{IEEEtran}
\bibliography{bibtex/bib/bibliog}

\section{Supplementary Materials}\label{AnexoI}

\begin{longtable}{llcccccc}
\caption{Regularized black box hybrid algorithms across different $p$ and $n$.
Note: RF= Random Forest; EN = Elastic Net} \\
\label{tab:hybrid_models_full} \\
\toprule
\textbf{Algorithm} & \textbf{Metric} & \textbf{$p$} & \textbf{$n=50$} & \textbf{$n=100$} & \textbf{$n=200$} & \textbf{$n=500$} & \textbf{$n=1000$} \\
\midrule
\endfirsthead

\multicolumn{8}{c}{{\bfseries \tablename\ \thetable{} -- continuation from previous page}} \\
\toprule
\textbf{Algorithm} & \textbf{Metric} & \textbf{$p$} & \textbf{$n=50$} & \textbf{$n=100$} & \textbf{$n=200$} & \textbf{$n=500$} & \textbf{$n=1000$} \\
\midrule
\endhead

\midrule \multicolumn{8}{r}{{Continued on next page}} \\
\endfoot

\bottomrule
\endlastfoot
CatBoost & Jaccard & 5 & 0.76 (0.08) & 0.8 (0.0) & 0.8 (0.0) & 0.8 (0.0) & 0.8 (0.0) \\
CatBoost & Jaccard & 10 & 0.65 (0.18) & 0.83 (0.14) & 0.89 (0.1) & 0.88 (0.13) & 0.95 (0.1) \\
CatBoost & Jaccard & 50 & 0.48 (0.13) & 0.77 (0.15) & 0.91 (0.12) & 0.95 (0.08) & 0.97 (0.07) \\
CatBoost & Jaccard & 100 & 0.31 (0.17) & 0.69 (0.15) & 0.8 (0.18) & 0.94 (0.1) & 1.0 (0.0) \\
CatBoost & RMSE & 5 & 2.97 (0.28) & 2.58 (0.16) & 2.32 (0.18) & 2.12 (0.09) & 2.01 (0.05) \\
CatBoost & RMSE & 10 & 2.94 (0.21) & 2.42 (0.11) & 1.94 (0.1) & 1.54 (0.06) & 1.34 (0.03) \\
CatBoost & RMSE & 50 & 3.15 (0.53) & 2.46 (0.22) & 1.87 (0.09) & 1.56 (0.04) & 1.35 (0.03) \\
CatBoost & RMSE & 100 & 3.4 (0.63) & 2.66 (0.3) & 1.96 (0.14) & 1.52 (0.06) & 1.36 (0.02) \\
CatBoost & Recovery & 5 & 0.76 (0.08) & 0.8 (0.0) & 0.8 (0.0) & 0.8 (0.0) & 0.8 (0.0) \\
CatBoost & Recovery & 10 & 0.74 (0.19) & 0.92 (0.1) & 1.0 (0.0) & 1.0 (0.0) & 1.0 (0.0) \\
CatBoost & Recovery & 50 & 0.64 (0.16) & 0.88 (0.1) & 1.0 (0.0) & 1.0 (0.0) & 1.0 (0.0) \\
CatBoost & Recovery & 100 & 0.5 (0.19) & 0.82 (0.18) & 0.98 (0.06) & 1.0 (0.0) & 1.0 (0.0) \\
CatBoost  EN & Jaccard & 5 & 0.86 (0.13) & 0.88 (0.1) & 0.86 (0.1) & 0.84 (0.08) & 0.84 (0.08) \\
CatBoost  EN & Jaccard & 10 & 0.62 (0.14) & 0.69 (0.11) & 0.67 (0.08) & 0.72 (0.1) & 0.72 (0.17) \\
CatBoost  EN & Jaccard & 50 & 0.47 (0.11) & 0.47 (0.11) & 0.45 (0.1) & 0.45 (0.08) & 0.46 (0.06) \\
CatBoost  EN & Jaccard & 100 & 0.4 (0.06) & 0.42 (0.09) & 0.44 (0.07) & 0.42 (0.08) & 0.44 (0.07) \\
CatBoost  EN & RMSE & 5 & 2.96 (0.23) & 2.58 (0.17) & 2.22 (0.28) & 1.99 (0.24) & 1.9 (0.32) \\
CatBoost  EN & RMSE & 10 & 3.27 (0.39) & 2.61 (0.21) & 2.41 (0.16) & 2.08 (0.26) & 1.73 (0.41) \\
CatBoost  EN & RMSE & 50 & 3.26 (0.35) & 2.81 (0.2) & 2.45 (0.28) & 2.22 (0.27) & 2.1 (0.22) \\
CatBoost  EN & RMSE & 100 & 3.4 (0.36) & 2.96 (0.1) & 2.53 (0.17) & 2.15 (0.27) & 1.97 (0.37) \\
CatBoost  EN & Recovery & 5 & 0.86 (0.13) & 0.88 (0.1) & 0.86 (0.1) & 0.84 (0.08) & 0.84 (0.08) \\
CatBoost  EN & Recovery & 10 & 0.9 (0.11) & 0.8 (0.0) & 0.82 (0.06) & 0.84 (0.08) & 0.9 (0.11) \\
CatBoost  EN & Recovery & 50 & 0.8 (0.09) & 0.8 (0.09) & 0.84 (0.08) & 0.84 (0.08) & 0.82 (0.06) \\
CatBoost  EN & Recovery & 100 & 0.78 (0.06) & 0.8 (0.0) & 0.8 (0.0) & 0.84 (0.08) & 0.86 (0.1) \\
CatBoost Lasso & Jaccard & 5 & 0.84 (0.16) & 0.82 (0.06) & 0.88 (0.1) & 0.88 (0.1) & 0.82 (0.06) \\
CatBoost Lasso & Jaccard & 10 & 0.68 (0.15) & 0.72 (0.16) & 0.65 (0.1) & 0.7 (0.12) & 0.68 (0.16) \\
CatBoost Lasso & Jaccard & 50 & 0.45 (0.11) & 0.42 (0.07) & 0.5 (0.13) & 0.45 (0.06) & 0.46 (0.06) \\
CatBoost Lasso & Jaccard & 100 & 0.41 (0.09) & 0.45 (0.07) & 0.46 (0.11) & 0.41 (0.05) & 0.49 (0.12) \\
CatBoost Lasso & RMSE & 5 & 3.06 (0.35) & 2.63 (0.18) & 2.18 (0.24) & 1.89 (0.34) & 1.99 (0.24) \\
CatBoost Lasso & RMSE & 10 & 3.15 (0.23) & 2.58 (0.23) & 2.42 (0.16) & 2.15 (0.2) & 1.82 (0.4) \\
CatBoost Lasso & RMSE & 50 & 3.31 (0.45) & 2.92 (0.15) & 2.46 (0.23) & 2.29 (0.2) & 2.18 (0.06) \\
CatBoost Lasso & RMSE & 100 & 3.35 (0.45) & 2.95 (0.1) & 2.43 (0.19) & 2.26 (0.21) & 1.88 (0.37) \\
CatBoost Lasso & Recovery & 5 & 0.84 (0.16) & 0.82 (0.06) & 0.88 (0.1) & 0.88 (0.1) & 0.82 (0.06) \\
CatBoost Lasso & Recovery & 10 & 0.88 (0.1) & 0.84 (0.08) & 0.82 (0.06) & 0.82 (0.06) & 0.88 (0.1) \\
CatBoost Lasso & Recovery & 50 & 0.78 (0.06) & 0.8 (0.0) & 0.84 (0.08) & 0.82 (0.06) & 0.8 (0.0) \\
CatBoost Lasso & Recovery & 100 & 0.74 (0.1) & 0.8 (0.0) & 0.82 (0.06) & 0.82 (0.06) & 0.88 (0.1) \\
CatBoost Ridge & Jaccard & 5 & 0.84 (0.08) & 0.82 (0.06) & 0.84 (0.08) & 0.84 (0.08) & 0.9 (0.11) \\
CatBoost Ridge & Jaccard & 10 & 0.65 (0.14) & 0.71 (0.15) & 0.64 (0.12) & 0.64 (0.15) & 0.69 (0.14) \\
CatBoost Ridge & Jaccard & 50 & 0.5 (0.15) & 0.43 (0.07) & 0.5 (0.14) & 0.44 (0.06) & 0.43 (0.07) \\
CatBoost Ridge & Jaccard & 100 & 0.43 (0.18) & 0.57 (0.17) & 0.45 (0.1) & 0.45 (0.09) & 0.47 (0.08) \\
CatBoost Ridge & RMSE & 5 & 2.98 (0.32) & 2.63 (0.18) & 2.23 (0.19) & 2.02 (0.26) & 1.69 (0.38) \\
CatBoost Ridge & RMSE & 10 & 3.17 (0.3) & 2.57 (0.2) & 2.41 (0.15) & 2.14 (0.2) & 1.81 (0.37) \\
CatBoost Ridge & RMSE & 50 & 3.28 (0.42) & 2.83 (0.2) & 2.44 (0.25) & 2.28 (0.23) & 2.09 (0.23) \\
CatBoost Ridge & RMSE & 100 & 3.45 (0.26) & 2.8 (0.14) & 2.54 (0.14) & 2.16 (0.24) & 1.95 (0.33) \\
CatBoost Ridge & Recovery & 5 & 0.84 (0.08) & 0.82 (0.06) & 0.84 (0.08) & 0.84 (0.08) & 0.9 (0.11) \\
CatBoost Ridge & Recovery & 10 & 0.88 (0.1) & 0.86 (0.1) & 0.84 (0.08) & 0.82 (0.06) & 0.88 (0.1) \\
CatBoost Ridge & Recovery & 50 & 0.74 (0.1) & 0.84 (0.08) & 0.84 (0.08) & 0.82 (0.06) & 0.82 (0.06) \\
CatBoost Ridge & Recovery & 100 & 0.64 (0.16) & 0.8 (0.0) & 0.8 (0.0) & 0.84 (0.08) & 0.86 (0.1) \\
ElasticNet & Jaccard & 5 & 0.86 (0.13) & 0.88 (0.1) & 0.86 (0.1) & 0.84 (0.08) & 0.84 (0.08) \\
ElasticNet & Jaccard & 10 & 0.62 (0.14) & 0.69 (0.11) & 0.67 (0.08) & 0.72 (0.1) & 0.72 (0.17) \\
ElasticNet & Jaccard & 50 & 0.47 (0.11) & 0.47 (0.11) & 0.45 (0.1) & 0.45 (0.08) & 0.46 (0.06) \\
ElasticNet & Jaccard & 100 & 0.4 (0.06) & 0.42 (0.09) & 0.44 (0.07) & 0.42 (0.08) & 0.44 (0.07) \\
ElasticNet & RMSE & 5 & 2.67 (0.36) & 2.7 (0.23) & 2.6 (0.24) & 2.65 (0.1) & 2.61 (0.06) \\
ElasticNet & RMSE & 10 & 2.42 (0.33) & 2.65 (0.22) & 2.61 (0.18) & 2.67 (0.1) & 2.65 (0.08) \\
ElasticNet & RMSE & 50 & 2.38 (0.58) & 2.6 (0.29) & 2.54 (0.18) & 2.64 (0.09) & 2.64 (0.05) \\
ElasticNet & RMSE & 100 & 2.41 (0.21) & 2.41 (0.13) & 2.46 (0.17) & 2.56 (0.16) & 2.6 (0.06) \\
ElasticNet & Recovery & 5 & 0.86 (0.13) & 0.88 (0.1) & 0.86 (0.1) & 0.84 (0.08) & 0.84 (0.08) \\
ElasticNet & Recovery & 10 & 0.9 (0.11) & 0.8 (0.0) & 0.82 (0.06) & 0.84 (0.08) & 0.9 (0.11) \\
ElasticNet & Recovery & 50 & 0.8 (0.09) & 0.8 (0.09) & 0.84 (0.08) & 0.84 (0.08) & 0.82 (0.06) \\
ElasticNet & Recovery & 100 & 0.78 (0.06) & 0.8 (0.0) & 0.8 (0.0) & 0.84 (0.08) & 0.86 (0.1) \\
H2OGBM & Jaccard & 5 & 0.76 (0.08) & 0.8 (0.0) & 0.8 (0.0) & 0.8 (0.0) & 0.8 (0.0) \\
H2OGBM & Jaccard & 10 & 0.67 (0.09) & 0.8 (0.16) & 0.9 (0.11) & 0.84 (0.16) & 0.83 (0.17) \\
H2OGBM & Jaccard & 50 & 0.49 (0.16) & 0.74 (0.2) & 0.89 (0.17) & 0.85 (0.16) & 0.86 (0.18) \\
H2OGBM & Jaccard & 100 & 0.42 (0.14) & 0.66 (0.19) & 0.92 (0.14) & 0.83 (0.18) & 0.91 (0.12) \\
H2OGBM & RMSE & 5 & 2.9 (0.24) & 2.63 (0.14) & 2.39 (0.15) & 2.21 (0.07) & 2.08 (0.08) \\
H2OGBM & RMSE & 10 & 2.75 (0.14) & 2.34 (0.17) & 2.06 (0.12) & 1.82 (0.04) & 1.6 (0.04) \\
H2OGBM & RMSE & 50 & 2.8 (0.34) & 2.41 (0.21) & 2.02 (0.1) & 1.8 (0.04) & 1.6 (0.03) \\
H2OGBM & RMSE & 100 & 2.79 (0.26) & 2.41 (0.17) & 2.05 (0.15) & 1.78 (0.07) & 1.6 (0.04) \\
H2OGBM & Recovery & 5 & 0.76 (0.08) & 0.8 (0.0) & 0.8 (0.0) & 0.8 (0.0) & 0.8 (0.0) \\
H2OGBM & Recovery & 10 & 0.86 (0.13) & 0.96 (0.08) & 1.0 (0.0) & 1.0 (0.0) & 1.0 (0.0) \\
H2OGBM & Recovery & 50 & 0.78 (0.15) & 0.94 (0.1) & 1.0 (0.0) & 1.0 (0.0) & 1.0 (0.0) \\
H2OGBM & Recovery & 100 & 0.74 (0.19) & 0.9 (0.11) & 1.0 (0.0) & 1.0 (0.0) & 1.0 (0.0) \\
H2OGBM  EN & Jaccard & 5 & 0.86 (0.13) & 0.88 (0.1) & 0.86 (0.1) & 0.84 (0.08) & 0.84 (0.08) \\
H2OGBM  EN & Jaccard & 10 & 0.62 (0.14) & 0.69 (0.11) & 0.67 (0.08) & 0.72 (0.1) & 0.72 (0.17) \\
H2OGBM  EN & Jaccard & 50 & 0.47 (0.11) & 0.47 (0.11) & 0.45 (0.1) & 0.45 (0.08) & 0.46 (0.06) \\
H2OGBM  EN & Jaccard & 100 & 0.4 (0.06) & 0.42 (0.09) & 0.44 (0.07) & 0.42 (0.08) & 0.44 (0.07) \\
H2OGBM  EN & RMSE & 5 & 3.0 (0.24) & 2.6 (0.19) & 2.31 (0.24) & 2.13 (0.19) & 1.98 (0.21) \\
H2OGBM  EN & RMSE & 10 & 3.0 (0.31) & 2.62 (0.18) & 2.47 (0.12) & 2.19 (0.21) & 1.86 (0.27) \\
H2OGBM  EN & RMSE & 50 & 3.07 (0.58) & 2.78 (0.3) & 2.43 (0.24) & 2.23 (0.17) & 2.09 (0.17) \\
H2OGBM  EN & RMSE & 100 & 3.23 (0.23) & 2.7 (0.14) & 2.46 (0.17) & 2.19 (0.23) & 1.99 (0.26) \\
H2OGBM  EN & Recovery & 5 & 0.86 (0.13) & 0.88 (0.1) & 0.86 (0.1) & 0.84 (0.08) & 0.84 (0.08) \\
H2OGBM  EN & Recovery & 10 & 0.9 (0.11) & 0.8 (0.0) & 0.82 (0.06) & 0.84 (0.08) & 0.9 (0.11) \\
H2OGBM  EN & Recovery & 50 & 0.8 (0.09) & 0.8 (0.09) & 0.84 (0.08) & 0.84 (0.08) & 0.82 (0.06) \\
H2OGBM  EN & Recovery & 100 & 0.78 (0.06) & 0.8 (0.0) & 0.8 (0.0) & 0.84 (0.08) & 0.86 (0.1) \\
H2OGBM Lasso & Jaccard & 5 & 0.84 (0.16) & 0.82 (0.06) & 0.88 (0.1) & 0.88 (0.1) & 0.82 (0.06) \\
H2OGBM Lasso & Jaccard & 10 & 0.68 (0.15) & 0.72 (0.16) & 0.65 (0.1) & 0.7 (0.12) & 0.68 (0.16) \\
H2OGBM Lasso & Jaccard & 50 & 0.45 (0.11) & 0.42 (0.07) & 0.5 (0.13) & 0.45 (0.06) & 0.46 (0.06) \\
H2OGBM Lasso & Jaccard & 100 & 0.41 (0.09) & 0.45 (0.07) & 0.46 (0.11) & 0.41 (0.05) & 0.49 (0.12) \\
H2OGBM Lasso & RMSE & 5 & 2.99 (0.24) & 2.68 (0.32) & 2.26 (0.18) & 2.04 (0.2) & 2.02 (0.15) \\
H2OGBM Lasso & RMSE & 10 & 3.0 (0.24) & 2.54 (0.24) & 2.45 (0.2) & 2.22 (0.17) & 1.91 (0.27) \\
H2OGBM Lasso & RMSE & 50 & 3.25 (0.45) & 2.75 (0.25) & 2.45 (0.24) & 2.28 (0.12) & 2.15 (0.06) \\
H2OGBM Lasso & RMSE & 100 & 3.25 (0.22) & 2.81 (0.16) & 2.43 (0.19) & 2.23 (0.16) & 1.95 (0.27) \\
H2OGBM Lasso & Recovery & 5 & 0.84 (0.16) & 0.82 (0.06) & 0.88 (0.1) & 0.88 (0.1) & 0.82 (0.06) \\
H2OGBM Lasso & Recovery & 10 & 0.88 (0.1) & 0.84 (0.08) & 0.82 (0.06) & 0.82 (0.06) & 0.88 (0.1) \\
H2OGBM Lasso & Recovery & 50 & 0.78 (0.06) & 0.8 (0.0) & 0.84 (0.08) & 0.82 (0.06) & 0.8 (0.0) \\
H2OGBM Lasso & Recovery & 100 & 0.74 (0.1) & 0.8 (0.0) & 0.82 (0.06) & 0.82 (0.06) & 0.88 (0.1) \\
H2OGBM Ridge & Jaccard & 5 & 0.84 (0.08) & 0.82 (0.06) & 0.84 (0.08) & 0.84 (0.08) & 0.9 (0.11) \\
H2OGBM Ridge & Jaccard & 10 & 0.65 (0.14) & 0.71 (0.15) & 0.64 (0.12) & 0.64 (0.15) & 0.69 (0.14) \\
H2OGBM Ridge & Jaccard & 50 & 0.5 (0.15) & 0.43 (0.07) & 0.5 (0.14) & 0.44 (0.06) & 0.43 (0.07) \\
H2OGBM Ridge & Jaccard & 100 & 0.43 (0.18) & 0.57 (0.17) & 0.45 (0.1) & 0.45 (0.09) & 0.47 (0.08) \\
H2OGBM Ridge & RMSE & 5 & 3.03 (0.28) & 2.6 (0.24) & 2.36 (0.19) & 2.15 (0.16) & 1.84 (0.24) \\
H2OGBM Ridge & RMSE & 10 & 3.13 (0.36) & 2.59 (0.25) & 2.43 (0.11) & 2.25 (0.16) & 1.9 (0.25) \\
H2OGBM Ridge & RMSE & 50 & 3.32 (0.62) & 2.74 (0.3) & 2.45 (0.25) & 2.28 (0.13) & 2.1 (0.17) \\
H2OGBM Ridge & RMSE & 100 & 3.41 (0.35) & 2.7 (0.14) & 2.48 (0.18) & 2.21 (0.23) & 1.98 (0.25) \\
H2OGBM Ridge & Recovery & 5 & 0.84 (0.08) & 0.82 (0.06) & 0.84 (0.08) & 0.84 (0.08) & 0.9 (0.11) \\
H2OGBM Ridge & Recovery & 10 & 0.88 (0.1) & 0.86 (0.1) & 0.84 (0.08) & 0.82 (0.06) & 0.88 (0.1) \\
H2OGBM Ridge & Recovery & 50 & 0.74 (0.1) & 0.84 (0.08) & 0.84 (0.08) & 0.82 (0.06) & 0.82 (0.06) \\
H2OGBM Ridge & Recovery & 100 & 0.64 (0.16) & 0.8 (0.0) & 0.8 (0.0) & 0.84 (0.08) & 0.86 (0.1) \\
Lasso & Jaccard & 5 & 0.84 (0.16) & 0.82 (0.06) & 0.88 (0.1) & 0.88 (0.1) & 0.82 (0.06) \\
Lasso & Jaccard & 10 & 0.68 (0.15) & 0.72 (0.16) & 0.65 (0.1) & 0.7 (0.12) & 0.68 (0.16) \\
Lasso & Jaccard & 50 & 0.45 (0.11) & 0.42 (0.07) & 0.5 (0.13) & 0.45 (0.06) & 0.46 (0.06) \\
Lasso & Jaccard & 100 & 0.41 (0.09) & 0.45 (0.07) & 0.46 (0.11) & 0.41 (0.05) & 0.49 (0.12) \\
Lasso & RMSE & 5 & 2.68 (0.35) & 2.68 (0.27) & 2.59 (0.24) & 2.65 (0.09) & 2.61 (0.06) \\
Lasso & RMSE & 10 & 2.4 (0.3) & 2.64 (0.24) & 2.61 (0.19) & 2.67 (0.11) & 2.65 (0.08) \\
Lasso & RMSE & 50 & 2.4 (0.55) & 2.61 (0.26) & 2.54 (0.17) & 2.65 (0.09) & 2.64 (0.05) \\
Lasso & RMSE & 100 & 2.41 (0.28) & 2.45 (0.13) & 2.45 (0.17) & 2.56 (0.16) & 2.6 (0.06) \\
Lasso & Recovery & 5 & 0.84 (0.16) & 0.82 (0.06) & 0.88 (0.1) & 0.88 (0.1) & 0.82 (0.06) \\
Lasso & Recovery & 10 & 0.88 (0.1) & 0.84 (0.08) & 0.82 (0.06) & 0.82 (0.06) & 0.88 (0.1) \\
Lasso & Recovery & 50 & 0.78 (0.06) & 0.8 (0.0) & 0.84 (0.08) & 0.82 (0.06) & 0.8 (0.0) \\
Lasso & Recovery & 100 & 0.74 (0.1) & 0.8 (0.0) & 0.82 (0.06) & 0.82 (0.06) & 0.88 (0.1) \\
LightGBM & Jaccard & 5 & 0.4 (0.21) & 0.8 (0.0) & 0.8 (0.0) & 0.8 (0.0) & 0.8 (0.0) \\
LightGBM & Jaccard & 10 & 0.54 (0.17) & 0.81 (0.12) & 0.8 (0.17) & 0.87 (0.1) & 0.91 (0.15) \\
LightGBM & Jaccard & 50 & 0.44 (0.12) & 0.76 (0.17) & 0.76 (0.18) & 0.76 (0.15) & 0.89 (0.14) \\
LightGBM & Jaccard & 100 & 0.41 (0.15) & 0.72 (0.2) & 0.76 (0.23) & 0.78 (0.19) & 0.95 (0.1) \\
LightGBM & RMSE & 5 & 5.12 (0.62) & 2.82 (0.14) & 2.4 (0.16) & 2.2 (0.07) & 2.07 (0.07) \\
LightGBM & RMSE & 10 & 5.07 (0.59) & 2.64 (0.11) & 2.1 (0.13) & 1.76 (0.04) & 1.55 (0.03) \\
LightGBM & RMSE & 50 & 4.88 (0.36) & 2.63 (0.24) & 2.04 (0.09) & 1.75 (0.05) & 1.56 (0.05) \\
LightGBM & RMSE & 100 & 5.04 (0.4) & 2.66 (0.13) & 2.09 (0.11) & 1.72 (0.08) & 1.56 (0.02) \\
LightGBM & Recovery & 5 & 0.4 (0.21) & 0.8 (0.0) & 0.8 (0.0) & 0.8 (0.0) & 0.8 (0.0) \\
LightGBM & Recovery & 10 & 0.7 (0.25) & 0.96 (0.08) & 1.0 (0.0) & 1.0 (0.0) & 1.0 (0.0) \\
LightGBM & Recovery & 50 & 0.68 (0.17) & 0.96 (0.08) & 1.0 (0.0) & 1.0 (0.0) & 1.0 (0.0) \\
LightGBM & Recovery & 100 & 0.56 (0.23) & 0.96 (0.08) & 1.0 (0.0) & 1.0 (0.0) & 1.0 (0.0) \\
LightGBM  EN & Jaccard & 5 & 0.86 (0.13) & 0.88 (0.1) & 0.86 (0.1) & 0.84 (0.08) & 0.84 (0.08) \\
LightGBM  EN & Jaccard & 10 & 0.62 (0.14) & 0.69 (0.11) & 0.67 (0.08) & 0.72 (0.1) & 0.72 (0.17) \\
LightGBM  EN & Jaccard & 50 & 0.47 (0.11) & 0.47 (0.11) & 0.45 (0.1) & 0.45 (0.08) & 0.46 (0.06) \\
LightGBM  EN & Jaccard & 100 & 0.4 (0.06) & 0.42 (0.09) & 0.44 (0.07) & 0.42 (0.08) & 0.44 (0.07) \\
LightGBM  EN & RMSE & 5 & 5.2 (0.64) & 2.82 (0.15) & 2.31 (0.25) & 2.08 (0.22) & 1.97 (0.23) \\
LightGBM  EN & RMSE & 10 & 5.25 (0.55) & 2.83 (0.14) & 2.44 (0.11) & 2.14 (0.2) & 1.85 (0.3) \\
LightGBM  EN & RMSE & 50 & 5.0 (0.32) & 2.92 (0.25) & 2.46 (0.24) & 2.21 (0.19) & 2.1 (0.18) \\
LightGBM  EN & RMSE & 100 & 5.16 (0.44) & 2.96 (0.15) & 2.5 (0.2) & 2.17 (0.23) & 2.0 (0.28) \\
LightGBM  EN & Recovery & 5 & 0.86 (0.13) & 0.88 (0.1) & 0.86 (0.1) & 0.84 (0.08) & 0.84 (0.08) \\
LightGBM  EN & Recovery & 10 & 0.9 (0.11) & 0.8 (0.0) & 0.82 (0.06) & 0.84 (0.08) & 0.9 (0.11) \\
LightGBM  EN & Recovery & 50 & 0.8 (0.09) & 0.8 (0.09) & 0.84 (0.08) & 0.84 (0.08) & 0.82 (0.06) \\
LightGBM  EN & Recovery & 100 & 0.78 (0.06) & 0.8 (0.0) & 0.8 (0.0) & 0.84 (0.08) & 0.86 (0.1) \\
LightGBM Lasso & Jaccard & 5 & 0.84 (0.16) & 0.82 (0.06) & 0.88 (0.1) & 0.88 (0.1) & 0.82 (0.06) \\
LightGBM Lasso & Jaccard & 10 & 0.68 (0.15) & 0.72 (0.16) & 0.65 (0.1) & 0.7 (0.12) & 0.68 (0.16) \\
LightGBM Lasso & Jaccard & 50 & 0.45 (0.11) & 0.42 (0.07) & 0.5 (0.13) & 0.45 (0.06) & 0.46 (0.06) \\
LightGBM Lasso & Jaccard & 100 & 0.41 (0.09) & 0.45 (0.07) & 0.46 (0.11) & 0.41 (0.05) & 0.49 (0.12) \\
LightGBM Lasso & RMSE & 5 & 5.2 (0.68) & 2.9 (0.17) & 2.25 (0.2) & 2.01 (0.23) & 2.02 (0.17) \\
LightGBM Lasso & RMSE & 10 & 5.22 (0.52) & 2.82 (0.12) & 2.43 (0.18) & 2.19 (0.17) & 1.9 (0.29) \\
LightGBM Lasso & RMSE & 50 & 5.0 (0.37) & 2.93 (0.27) & 2.44 (0.25) & 2.27 (0.17) & 2.16 (0.07) \\
LightGBM Lasso & RMSE & 100 & 5.22 (0.39) & 2.96 (0.14) & 2.48 (0.17) & 2.22 (0.18) & 1.94 (0.28) \\
LightGBM Lasso & Recovery & 5 & 0.84 (0.16) & 0.82 (0.06) & 0.88 (0.1) & 0.88 (0.1) & 0.82 (0.06) \\
LightGBM Lasso & Recovery & 10 & 0.88 (0.1) & 0.84 (0.08) & 0.82 (0.06) & 0.82 (0.06) & 0.88 (0.1) \\
LightGBM Lasso & Recovery & 50 & 0.78 (0.06) & 0.8 (0.0) & 0.84 (0.08) & 0.82 (0.06) & 0.8 (0.0) \\
LightGBM Lasso & Recovery & 100 & 0.74 (0.1) & 0.8 (0.0) & 0.82 (0.06) & 0.82 (0.06) & 0.88 (0.1) \\
LightGBM Ridge & Jaccard & 5 & 0.84 (0.08) & 0.82 (0.06) & 0.84 (0.08) & 0.84 (0.08) & 0.9 (0.11) \\
LightGBM Ridge & Jaccard & 10 & 0.65 (0.14) & 0.71 (0.15) & 0.64 (0.12) & 0.64 (0.15) & 0.69 (0.14) \\
LightGBM Ridge & Jaccard & 50 & 0.5 (0.15) & 0.43 (0.07) & 0.5 (0.14) & 0.44 (0.06) & 0.43 (0.07) \\
LightGBM Ridge & Jaccard & 100 & 0.43 (0.18) & 0.57 (0.17) & 0.45 (0.1) & 0.45 (0.09) & 0.47 (0.08) \\
LightGBM Ridge & RMSE & 5 & 5.26 (0.63) & 2.8 (0.13) & 2.34 (0.15) & 2.1 (0.14) & 1.8 (0.25) \\
LightGBM Ridge & RMSE & 10 & 5.17 (0.57) & 2.79 (0.17) & 2.43 (0.1) & 2.22 (0.15) & 1.89 (0.29) \\
LightGBM Ridge & RMSE & 50 & 5.04 (0.39) & 2.96 (0.25) & 2.45 (0.25) & 2.26 (0.13) & 2.1 (0.17) \\
LightGBM Ridge & RMSE & 100 & 5.11 (0.39) & 2.94 (0.12) & 2.48 (0.19) & 2.15 (0.21) & 1.99 (0.27) \\
LightGBM Ridge & Recovery & 5 & 0.84 (0.08) & 0.82 (0.06) & 0.84 (0.08) & 0.84 (0.08) & 0.9 (0.11) \\
LightGBM Ridge & Recovery & 10 & 0.88 (0.1) & 0.86 (0.1) & 0.84 (0.08) & 0.82 (0.06) & 0.88 (0.1) \\
LightGBM Ridge & Recovery & 50 & 0.74 (0.1) & 0.84 (0.08) & 0.84 (0.08) & 0.82 (0.06) & 0.82 (0.06) \\
LightGBM Ridge & Recovery & 100 & 0.64 (0.16) & 0.8 (0.0) & 0.8 (0.0) & 0.84 (0.08) & 0.86 (0.1) \\
RF & Jaccard & 5 & 0.76 (0.08) & 0.74 (0.1) & 0.78 (0.06) & 0.8 (0.0) & 0.8 (0.0) \\
RF & Jaccard & 10 & 0.71 (0.17) & 0.83 (0.15) & 0.81 (0.09) & 0.83 (0.0) & 0.83 (0.0) \\
RF & Jaccard & 50 & 0.63 (0.13) & 0.72 (0.18) & 0.83 (0.0) & 0.83 (0.0) & 0.83 (0.0) \\
RF & Jaccard & 100 & 0.49 (0.12) & 0.72 (0.12) & 0.78 (0.09) & 0.83 (0.0) & 0.83 (0.0) \\
RF & RMSE & 5 & 3.26 (0.32) & 2.87 (0.12) & 2.62 (0.12) & 2.4 (0.05) & 2.23 (0.05) \\
RF & RMSE & 10 & 3.07 (0.22) & 2.81 (0.1) & 2.51 (0.08) & 2.15 (0.07) & 1.91 (0.04) \\
RF & RMSE & 50 & 3.02 (0.32) & 2.87 (0.16) & 2.44 (0.08) & 2.16 (0.05) & 1.92 (0.05) \\
RF & RMSE & 100 & 3.26 (0.35) & 2.86 (0.13) & 2.5 (0.16) & 2.12 (0.09) & 1.9 (0.03) \\
RF & Recovery & 5 & 0.76 (0.08) & 0.74 (0.1) & 0.78 (0.06) & 0.8 (0.0) & 0.8 (0.0) \\
RF & Recovery & 10 & 0.8 (0.21) & 0.88 (0.17) & 1.0 (0.0) & 1.0 (0.0) & 1.0 (0.0) \\
RF & Recovery & 50 & 0.68 (0.14) & 0.86 (0.16) & 1.0 (0.0) & 1.0 (0.0) & 1.0 (0.0) \\
RF & Recovery & 100 & 0.58 (0.15) & 0.8 (0.13) & 0.98 (0.06) & 1.0 (0.0) & 1.0 (0.0) \\
RF  EN & Jaccard & 5 & 0.86 (0.13) & 0.88 (0.1) & 0.86 (0.1) & 0.84 (0.08) & 0.84 (0.08) \\
RF  EN & Jaccard & 10 & 0.62 (0.14) & 0.69 (0.11) & 0.67 (0.08) & 0.72 (0.1) & 0.72 (0.17) \\
RF  EN & Jaccard & 50 & 0.47 (0.11) & 0.47 (0.11) & 0.45 (0.1) & 0.45 (0.08) & 0.46 (0.06) \\
RF  EN & Jaccard & 100 & 0.4 (0.06) & 0.42 (0.09) & 0.44 (0.07) & 0.42 (0.08) & 0.44 (0.07) \\
RF  EN & RMSE & 5 & 3.33 (0.34) & 2.9 (0.1) & 2.6 (0.17) & 2.36 (0.09) & 2.18 (0.11) \\
RF  EN & RMSE & 10 & 3.45 (0.23) & 2.95 (0.14) & 2.78 (0.12) & 2.43 (0.19) & 2.16 (0.19) \\
RF  EN & RMSE & 50 & 3.41 (0.34) & 3.11 (0.17) & 2.78 (0.12) & 2.53 (0.12) & 2.37 (0.09) \\
RF  EN & RMSE & 100 & 3.55 (0.29) & 3.18 (0.11) & 2.78 (0.18) & 2.48 (0.14) & 2.28 (0.16) \\
RF  EN & Recovery & 5 & 0.86 (0.13) & 0.88 (0.1) & 0.86 (0.1) & 0.84 (0.08) & 0.84 (0.08) \\
RF  EN & Recovery & 10 & 0.9 (0.11) & 0.8 (0.0) & 0.82 (0.06) & 0.84 (0.08) & 0.9 (0.11) \\
RF  EN & Recovery & 50 & 0.8 (0.09) & 0.8 (0.09) & 0.84 (0.08) & 0.84 (0.08) & 0.82 (0.06) \\
RF  EN & Recovery & 100 & 0.78 (0.06) & 0.8 (0.0) & 0.8 (0.0) & 0.84 (0.08) & 0.86 (0.1) \\
RF Lasso & Jaccard & 5 & 0.84 (0.16) & 0.82 (0.06) & 0.88 (0.1) & 0.88 (0.1) & 0.82 (0.06) \\
RF Lasso & Jaccard & 10 & 0.68 (0.15) & 0.72 (0.16) & 0.65 (0.1) & 0.7 (0.12) & 0.68 (0.16) \\
RF Lasso & Jaccard & 50 & 0.45 (0.11) & 0.42 (0.07) & 0.5 (0.13) & 0.45 (0.06) & 0.46 (0.06) \\
RF Lasso & Jaccard & 100 & 0.41 (0.09) & 0.45 (0.07) & 0.46 (0.11) & 0.41 (0.05) & 0.49 (0.12) \\
RF Lasso & RMSE & 5 & 3.32 (0.27) & 2.93 (0.12) & 2.59 (0.12) & 2.33 (0.08) & 2.21 (0.08) \\
RF Lasso & RMSE & 10 & 3.37 (0.17) & 2.94 (0.14) & 2.75 (0.12) & 2.47 (0.09) & 2.2 (0.21) \\
RF Lasso & RMSE & 50 & 3.38 (0.34) & 3.18 (0.17) & 2.75 (0.17) & 2.56 (0.11) & 2.37 (0.06) \\
RF Lasso & RMSE & 100 & 3.47 (0.37) & 3.17 (0.19) & 2.82 (0.15) & 2.51 (0.11) & 2.25 (0.16) \\
RF Lasso & Recovery & 5 & 0.84 (0.16) & 0.82 (0.06) & 0.88 (0.1) & 0.88 (0.1) & 0.82 (0.06) \\
RF Lasso & Recovery & 10 & 0.88 (0.1) & 0.84 (0.08) & 0.82 (0.06) & 0.82 (0.06) & 0.88 (0.1) \\
RF Lasso & Recovery & 50 & 0.78 (0.06) & 0.8 (0.0) & 0.84 (0.08) & 0.82 (0.06) & 0.8 (0.0) \\
RF Lasso & Recovery & 100 & 0.74 (0.1) & 0.8 (0.0) & 0.82 (0.06) & 0.82 (0.06) & 0.88 (0.1) \\
RF Ridge & Jaccard & 5 & 0.84 (0.08) & 0.82 (0.06) & 0.84 (0.08) & 0.84 (0.08) & 0.9 (0.11) \\
RF Ridge & Jaccard & 10 & 0.65 (0.14) & 0.71 (0.15) & 0.64 (0.12) & 0.64 (0.15) & 0.69 (0.14) \\
RF Ridge & Jaccard & 50 & 0.5 (0.15) & 0.43 (0.07) & 0.5 (0.14) & 0.44 (0.06) & 0.43 (0.07) \\
RF Ridge & Jaccard & 100 & 0.43 (0.18) & 0.57 (0.17) & 0.45 (0.1) & 0.45 (0.09) & 0.47 (0.08) \\
RF Ridge & RMSE & 5 & 3.35 (0.31) & 2.9 (0.09) & 2.61 (0.1) & 2.37 (0.07) & 2.12 (0.11) \\
RF Ridge & RMSE & 10 & 3.38 (0.22) & 2.95 (0.11) & 2.74 (0.12) & 2.48 (0.1) & 2.21 (0.15) \\
RF Ridge & RMSE & 50 & 3.42 (0.4) & 3.16 (0.23) & 2.76 (0.14) & 2.53 (0.12) & 2.35 (0.12) \\
RF Ridge & RMSE & 100 & 3.54 (0.38) & 3.17 (0.07) & 2.8 (0.2) & 2.5 (0.15) & 2.28 (0.16) \\
RF Ridge & Recovery & 5 & 0.84 (0.08) & 0.82 (0.06) & 0.84 (0.08) & 0.84 (0.08) & 0.9 (0.11) \\
RF Ridge & Recovery & 10 & 0.88 (0.1) & 0.86 (0.1) & 0.84 (0.08) & 0.82 (0.06) & 0.88 (0.1) \\
RF Ridge & Recovery & 50 & 0.74 (0.1) & 0.84 (0.08) & 0.84 (0.08) & 0.82 (0.06) & 0.82 (0.06) \\
RF Ridge & Recovery & 100 & 0.64 (0.16) & 0.8 (0.0) & 0.8 (0.0) & 0.84 (0.08) & 0.86 (0.1) \\
Ridge & Jaccard & 5 & 0.84 (0.08) & 0.82 (0.06) & 0.84 (0.08) & 0.84 (0.08) & 0.9 (0.11) \\
Ridge & Jaccard & 10 & 0.65 (0.14) & 0.71 (0.15) & 0.64 (0.12) & 0.64 (0.15) & 0.69 (0.14) \\
Ridge & Jaccard & 50 & 0.5 (0.15) & 0.43 (0.07) & 0.5 (0.14) & 0.44 (0.06) & 0.43 (0.07) \\
Ridge & Jaccard & 100 & 0.43 (0.18) & 0.57 (0.17) & 0.45 (0.1) & 0.45 (0.09) & 0.47 (0.08) \\
Ridge & RMSE & 5 & 2.67 (0.39) & 2.71 (0.24) & 2.6 (0.25) & 2.65 (0.1) & 2.61 (0.06) \\
Ridge & RMSE & 10 & 2.41 (0.35) & 2.64 (0.23) & 2.62 (0.17) & 2.67 (0.1) & 2.65 (0.08) \\
Ridge & RMSE & 50 & 2.65 (0.54) & 2.62 (0.29) & 2.54 (0.17) & 2.65 (0.09) & 2.65 (0.05) \\
Ridge & RMSE & 100 & 3.01 (0.33) & 2.57 (0.15) & 2.49 (0.18) & 2.57 (0.15) & 2.6 (0.06) \\
Ridge & Recovery & 5 & 0.84 (0.08) & 0.82 (0.06) & 0.84 (0.08) & 0.84 (0.08) & 0.9 (0.11) \\
Ridge & Recovery & 10 & 0.88 (0.1) & 0.86 (0.1) & 0.84 (0.08) & 0.82 (0.06) & 0.88 (0.1) \\
Ridge & Recovery & 50 & 0.74 (0.1) & 0.84 (0.08) & 0.84 (0.08) & 0.82 (0.06) & 0.82 (0.06) \\
Ridge & Recovery & 100 & 0.64 (0.16) & 0.8 (0.0) & 0.8 (0.0) & 0.84 (0.08) & 0.86 (0.1) \\
XGBoost & Jaccard & 5 & 0.76 (0.08) & 0.74 (0.1) & 0.8 (0.0) & 0.8 (0.0) & 0.8 (0.0) \\
XGBoost & Jaccard & 10 & 0.68 (0.17) & 0.78 (0.18) & 0.9 (0.14) & 0.92 (0.11) & 0.94 (0.1) \\
XGBoost & Jaccard & 50 & 0.51 (0.15) & 0.62 (0.21) & 0.91 (0.13) & 0.88 (0.15) & 0.97 (0.07) \\
XGBoost & Jaccard & 100 & 0.41 (0.17) & 0.54 (0.17) & 0.84 (0.12) & 0.93 (0.09) & 0.98 (0.05) \\
XGBoost & RMSE & 5 & 3.06 (0.37) & 2.8 (0.17) & 2.57 (0.19) & 2.35 (0.05) & 2.2 (0.06) \\
XGBoost & RMSE & 10 & 2.88 (0.2) & 2.52 (0.15) & 2.27 (0.09) & 1.91 (0.04) & 1.68 (0.04) \\
XGBoost & RMSE & 50 & 3.03 (0.43) & 2.7 (0.29) & 2.24 (0.16) & 1.93 (0.06) & 1.69 (0.05) \\
XGBoost & RMSE & 100 & 3.02 (0.37) & 2.73 (0.23) & 2.25 (0.14) & 1.91 (0.06) & 1.67 (0.04) \\
XGBoost & Recovery & 5 & 0.76 (0.08) & 0.74 (0.1) & 0.8 (0.0) & 0.8 (0.0) & 0.8 (0.0) \\
XGBoost & Recovery & 10 & 0.88 (0.14) & 0.94 (0.1) & 1.0 (0.0) & 1.0 (0.0) & 1.0 (0.0) \\
XGBoost & Recovery & 50 & 0.68 (0.17) & 0.88 (0.1) & 0.98 (0.06) & 1.0 (0.0) & 1.0 (0.0) \\
XGBoost & Recovery & 100 & 0.66 (0.19) & 0.84 (0.08) & 1.0 (0.0) & 1.0 (0.0) & 1.0 (0.0) \\
XGBoost  EN & Jaccard & 5 & 0.86 (0.13) & 0.88 (0.1) & 0.86 (0.1) & 0.84 (0.08) & 0.84 (0.08) \\
XGBoost  EN & Jaccard & 10 & 0.62 (0.14) & 0.69 (0.11) & 0.67 (0.08) & 0.72 (0.1) & 0.72 (0.17) \\
XGBoost  EN & Jaccard & 50 & 0.47 (0.11) & 0.47 (0.11) & 0.45 (0.1) & 0.45 (0.08) & 0.46 (0.06) \\
XGBoost  EN & Jaccard & 100 & 0.4 (0.06) & 0.42 (0.09) & 0.44 (0.07) & 0.42 (0.08) & 0.44 (0.07) \\
XGBoost  EN & RMSE & 5 & 3.15 (0.31) & 2.8 (0.17) & 2.49 (0.31) & 2.27 (0.25) & 2.11 (0.25) \\
XGBoost  EN & RMSE & 10 & 3.27 (0.41) & 2.81 (0.31) & 2.63 (0.14) & 2.32 (0.2) & 1.98 (0.3) \\
XGBoost  EN & RMSE & 50 & 3.33 (0.46) & 2.89 (0.27) & 2.67 (0.23) & 2.43 (0.15) & 2.28 (0.15) \\
XGBoost  EN & RMSE & 100 & 3.43 (0.48) & 3.03 (0.17) & 2.66 (0.22) & 2.37 (0.21) & 2.15 (0.24) \\
XGBoost  EN & Recovery & 5 & 0.86 (0.13) & 0.88 (0.1) & 0.86 (0.1) & 0.84 (0.08) & 0.84 (0.08) \\
XGBoost  EN & Recovery & 10 & 0.9 (0.11) & 0.8 (0.0) & 0.82 (0.06) & 0.84 (0.08) & 0.9 (0.11) \\
XGBoost  EN & Recovery & 50 & 0.8 (0.09) & 0.8 (0.09) & 0.84 (0.08) & 0.84 (0.08) & 0.82 (0.06) \\
XGBoost  EN & Recovery & 100 & 0.78 (0.06) & 0.8 (0.0) & 0.8 (0.0) & 0.84 (0.08) & 0.86 (0.1) \\
XGBoost Lasso & Jaccard & 5 & 0.84 (0.16) & 0.82 (0.06) & 0.88 (0.1) & 0.88 (0.1) & 0.82 (0.06) \\
XGBoost Lasso & Jaccard & 10 & 0.68 (0.15) & 0.72 (0.16) & 0.65 (0.1) & 0.7 (0.12) & 0.68 (0.16) \\
XGBoost Lasso & Jaccard & 50 & 0.45 (0.11) & 0.42 (0.07) & 0.5 (0.13) & 0.45 (0.06) & 0.46 (0.06) \\
XGBoost Lasso & Jaccard & 100 & 0.41 (0.09) & 0.45 (0.07) & 0.46 (0.11) & 0.41 (0.05) & 0.49 (0.12) \\
XGBoost Lasso & RMSE & 5 & 3.17 (0.25) & 2.75 (0.24) & 2.49 (0.21) & 2.17 (0.24) & 2.16 (0.17) \\
XGBoost Lasso & RMSE & 10 & 3.27 (0.48) & 2.72 (0.19) & 2.58 (0.24) & 2.4 (0.15) & 2.04 (0.29) \\
XGBoost Lasso & RMSE & 50 & 3.3 (0.49) & 3.01 (0.26) & 2.65 (0.15) & 2.45 (0.13) & 2.32 (0.09) \\
XGBoost Lasso & RMSE & 100 & 3.29 (0.43) & 3.01 (0.19) & 2.64 (0.21) & 2.39 (0.14) & 2.1 (0.28) \\
XGBoost Lasso & Recovery & 5 & 0.84 (0.16) & 0.82 (0.06) & 0.88 (0.1) & 0.88 (0.1) & 0.82 (0.06) \\
XGBoost Lasso & Recovery & 10 & 0.88 (0.1) & 0.84 (0.08) & 0.82 (0.06) & 0.82 (0.06) & 0.88 (0.1) \\
XGBoost Lasso & Recovery & 50 & 0.78 (0.06) & 0.8 (0.0) & 0.84 (0.08) & 0.82 (0.06) & 0.8 (0.0) \\
XGBoost Lasso & Recovery & 100 & 0.74 (0.1) & 0.8 (0.0) & 0.82 (0.06) & 0.82 (0.06) & 0.88 (0.1) \\
XGBoost Ridge & Jaccard & 5 & 0.84 (0.08) & 0.82 (0.06) & 0.84 (0.08) & 0.84 (0.08) & 0.9 (0.11) \\
XGBoost Ridge & Jaccard & 10 & 0.65 (0.14) & 0.71 (0.15) & 0.64 (0.12) & 0.64 (0.15) & 0.69 (0.14) \\
XGBoost Ridge & Jaccard & 50 & 0.5 (0.15) & 0.43 (0.07) & 0.5 (0.14) & 0.44 (0.06) & 0.43 (0.07) \\
XGBoost Ridge & Jaccard & 100 & 0.43 (0.18) & 0.57 (0.17) & 0.45 (0.1) & 0.45 (0.09) & 0.47 (0.08) \\
XGBoost Ridge & RMSE & 5 & 3.18 (0.3) & 2.82 (0.24) & 2.53 (0.23) & 2.28 (0.14) & 1.93 (0.27) \\
XGBoost Ridge & RMSE & 10 & 3.22 (0.29) & 2.74 (0.16) & 2.58 (0.13) & 2.36 (0.14) & 2.05 (0.26) \\
XGBoost Ridge & RMSE & 50 & 3.36 (0.54) & 2.95 (0.27) & 2.67 (0.2) & 2.47 (0.12) & 2.27 (0.17) \\
XGBoost Ridge & RMSE & 100 & 3.55 (0.67) & 2.98 (0.13) & 2.67 (0.16) & 2.38 (0.21) & 2.16 (0.25) \\
XGBoost Ridge & Recovery & 5 & 0.84 (0.08) & 0.82 (0.06) & 0.84 (0.08) & 0.84 (0.08) & 0.9 (0.11) \\
XGBoost Ridge & Recovery & 10 & 0.88 (0.1) & 0.86 (0.1) & 0.84 (0.08) & 0.82 (0.06) & 0.88 (0.1) \\
XGBoost Ridge & Recovery & 50 & 0.74 (0.1) & 0.84 (0.08) & 0.84 (0.08) & 0.82 (0.06) & 0.82 (0.06) \\
XGBoost Ridge & Recovery & 100 & 0.64 (0.16) & 0.8 (0.0) & 0.8 (0.0) & 0.84 (0.08) & 0.86 (0.1) \\
\end{longtable}

\end{document}